# Measuring University Impact: Wikipedia approach

Tatiana Kozitsina (Babkina)[1,2]*, Viacheslav Goiko[3], Roman Palkin[3], Valentin Khomutenko[3], Yulia Mundrievskaya[3], Maria Sukhareva[4], Isak Froumin[5], and Mikhail Myagkov[3,5,6]

[1]Moscow Institute of Physics and Technology (State University), Department of Control and Applied Mathematics, 9 Institutskiy per., Dolgoprudny, Moscow Region, 141701, Russian Federation.

[2]Federal Research Center Computer Science and Control, Russian Academy of Sciences, Moscow, 119333, Russian Federation

[3]Tomsk State University, Laboratory of Big Data in Social Sciences, 36 Lenin Ave., Tomsk, 634050, Russian Federation.

[4]Moscow State University, Public administration faculty, 27(4) Lomonosov Pr., Moscow, 119991, Russian Federation.

[5]Institute of Education, National Research University Higher School of Economics, 16 Potpovskiy Pereulok, building 10, Moscow, 101000, Russian Federation.

[6]Department of Political Science, University of Oregon, Oregon, 97403, United States.

* Corresponding author (e-mail: babkinats@yandex.ru)

## Abstract

The impact of Universities on the social, economic and political landscape is one of the key directions in contemporary educational evaluation. In this paper, we discuss the new methodological technique that evaluates the impact of university based on popularity (number of page-views) of their alumni's pages on Wikipedia. It allows revealing the alumni popularity dynamics and tracking its state. Preliminary analysis shows that the number of page-views is higher for the contemporary persons that prove the perspectives of this approach. Then, universities were ranked based on the methodology and compared to the famous international university rankings ARWU and QS based only on alumni scales: for the top 10 universities, there is an intersection of two universities (Columbia University, Stanford University). The correlation coefficients between different university rankings are provided in the paper. Finally, the ranking based on the alumni popularity was compared with the ranking of universities based on the popularity of their webpages on Wikipedia: there is a strong connection between these indicators.

## Introduction

As a compulsory component of the progress engine, university is an important organization, which affects many strata of a society. University is not only about education, but also about social connections, career mobility, scientific discoveries and achievements, student movements, and alumni unions. So, firstly, university is defined through the people: students, teachers, scientists, graduates, etc. Secondly, it can be noticed, that a university is a forge of thoughts; however, humanity has not yet managed to comprehend the measurement of thinking. And while there is no doubt about university's significance, there appears a question of how to evaluate university's influence, which seems to be a big and ambitious goal (Agasisti 2017; Khazragui and Hudson 2014; Witte and López-Torres 2017).

By this moment, the directions of university's activities can be divided into three components: education, science, and communication with the society – university's "third mission" (Laredo 2007; Lukman et al. 2010; Montesinos et al. 2008; Scott 2006). The influence of a university may be measured through the first two components by assessing the quality of admission, students' grades, quantity of articles and citations, number of awards won, grants received (Brooks 2005). But the role of the university's contribution to society is a crucial issue which is common debates among researchers (Fumasoli 2016; Kosmützky and Ewen 2016). University's impact on society can be assessed through the success of graduates, joint projects of the university with external companies, transfer of knowledge to society, usage of university's research by society, and interaction between university and the region (Olcay and Bulu 2017;

Sánchez-Barrioluengo 2014; Secundo et al. 2017; Uyarra 2010). Such measurements are not entirely obvious and require the development of new methods. In relation to this, the growing rates of life digitalization may become the key to discover university's impact on society.

The impact of a modern university largely depends on its presence on the online information sites. Digital data from Internet resources can be used to form new metrics for evaluating university activities. In recent years, altimetric has emerged in the scientific community - non-traditional ways of evaluating scientific activities based on digital traces - mentions in social media, count of downloads, views, etc. (Altmetric 2019; Chavda and Patel 2016; Nariani 2017). By analogy, this way we can approach the task of evaluating universities. For example, the influence of a university directly depends on the influence and success of its graduates. However, data on the success of the graduate can be obtained from Wikipedia, social networks (Kempe et al. 2005; Kozitsin 2020; Kozitsin et al. 2020) and other information online sites. In this paper, we propose to use the data on graduates from a multilingual, web-based, free encyclopedia Wikipedia (Wikipedia 2019) to analyze their impact on society. We suggest using the number of page-views of the graduates' webpages on Wikipedia as a metrics for universities comparison.

Alumnus is the results of the universities' production activities. Their career success represents the quality of the educational programs and the university's system (Vermeulen and Schmidt 2008). Graduates involuntarily become the face of the university and in some cases a potential source of income (Scillitoe 2013). Commonly, only famous graduates get into consideration for the research purposes: people winning Nobel Prizes and Fields Medals (Shanghai Ranking 2017), or graduates who stay in science, or higher education (Top Universities 2019). At the same time, the significance of the university is standing through the influence of the graduates. For that purpose, we should learn how to locate all the important for society graduates and how to estimate their influence.

Traditional studies of the alumni use the survey method, for example, questionnaires sent to the graduates and employers (Top Universities 2019). The main focus of such studies is the work satisfaction (Laschinger 2012), meeting expectations (Nicolescu & Paun 2009), the level of salary (Lee and Wilbur 1985), employer's estimation (Cole and Thompson 2002), competencies and skills obtained from higher education (Hernández-March et al. 2009). Cabrera et al. (2005) show that from the 1930s to 1970s researchers studied the satisfaction from work, a connection between education, and the employment field. After the 1980s the focus of studies shifted to the competencies received at the university, and also alumni engagement in the university life. It was found (Ng et al. 2005; Seibert et al. 1999) which psychological characteristics are connected with career success. Additionally, career success relates positively to the internships during the education processes (Gault et al. 2000). Also, the research focused on the period of time which graduate spent to find the job (Quarmby et al. 1999).

Despite the fact that the survey method stays actual and useful, scientists started to pay attention to the data from the Web to study alumni (Ranking Web of Universities 2019). In the paper (Gonçalves et al. 2014) the information from the business social network LinkedIn (LinkedIn 2019) was gathered. The uniqueness of this approach is that the users of this site post there a lot of information about their education, work experience, expectations — the important characteristics of the alumni investigation. Papoutsoglou et al. (2017) proposed a similar method. They created the algorithm to information crawling from the Stack Overflow Job Advertisements (Stack Overflow 2019). The separate way of the alumni research is to study the social connections (Nann et al. 2010; Rubens et al. 2011). The main conclusion here is that the wide social network facilitates the career success.

Besides the scientific literature, we studied the methods of the international university rankings. We found that alumni success takes a little role in forming the rankings: QS (Employer Reputation — 10%) (Top Universities 2019), Times Higher Education (0%) (The World University Rankings 2019), ARWU (Alumni of an institution winning Nobel Prizes and Fields Medals — 10%) (Shanghai Ranking 2017). The remaining indicators in the rankings refer to science and students. The general issue here is that searching for information about alumni is hard to conduct automatically. And the way to extract the information about alumni from Wikipedia solve this problem with automatization.

It should be noticed that all persons must go through the "Wikipedia Notability for people" (Wikipedia: Notability (people) 2019) to have the own webpage on Wikipedia. Thus, we can find on Wikipedia only those people who made some significant impact. According to the Wikipedia rules, a person should be "remarkable" or "significant, interesting, or unusual enough to deserve attention or to be recorded" (Wikipedia: Notability (people) 2019). Being famous or popular is not the primary criteria for dedicating personal page. A person must be mentioned on the other sources (independent from this person), or must have a well-known and significant award or honor, or must have some historical importance, or must be the scholars with famous and important ideas, etc. Moreover, the relationship with the famous person is not enough to get a personal page on Wikipedia. Finding information about university graduates on Wikipedia allows not only to cover the most important graduates, but also to understand how important these graduates are to society. The number of page-views of the graduates' webpages shows the popularity, familiarity, and interest.

The research goal of the paper is to discuss the possibility to evaluate the university impact based on the importance of its graduates to society and to compare the result with the well-known world university rankings. The methodological goal of the research is to demonstrate the method of collecting information on alumni through the number of page-views of their webpages on the free encyclopedia Wikipedia. For this purpose, we created a database containing 265,709 graduates from 464 inter-national universities. For every alumnus we found their name in English and national language, university (alma mater), birth year, link to the webpages in English and national languages, the total sum of views from webpages in English and national languages. Final sample contains graduates born from the X century to the XX century. The results of the investigation of the dataset are proposed for consideration and compared with the existing university rankings. The perspective of using the methods is discussed.

# Methods

Firstly, we defined the list of top 464 universities. Then we created the database which contained all the webpages of the chosen universities on Wikipedia (in English and national language). Finally, we identified all the webpages devoted to people who were graduated from the chosen university (in English and national language). After that, we found the total amount of views of these found webpages devoted to graduates and universities for 2017 year. The full procedure is described in Appendix.

The final dataset includes: 265,709 graduates from 464 Universities and the sum of views of their Wikipedia pages; 464 Universities and the sum of views of their Wikipedia pages. The following parameters were taken into consideration for the current research:

- The sum of graduates from the particular University;
- The birth year of the graduates from the particular University;
- The sum of views of the graduate's Wikipedia page;
- The sum of views of the university's Wikipedia page;
- The total sum of the all graduate's webpage views for the particular University.

Thus, current results are based on the statistics about number of webpage-views of the alumni and universities on Wikipedia. For comparison, we observed three popular universities rankings for 2017 year respectively: QS World University Rankings (later - QS) (Top Universities 2017), QS Graduate Employability Rankings (QS Graduate Employability Rankings 2017), Academic Ranking of World Universities (later - ARWU) (Shanghai Ranking 2017). These rankings were chosen because they took into consideration data about alumni.

Further, we provide a review of these rankings.

ARWU is formed on the basis of 6 indicators: alumni of an institution winning Nobel Prizes and Fields Medals, staff of an institution winning Nobel Prizes and Fields Medals, highly cited researchers in 21 broad subject categories, papers published in Nature and Science, papers indexed in Science Citation Index-expanded and Social Science Citation Index, per capita academic performance of an institution. Alumni indicator is based on those who obtained bachelor's, master's or doctoral degrees from the university. The weight for particular alumnus is depended on the year of graduating: the biggest (100%) for 2001-2010, the lowest (10%) for 1911-1920. Thus, this indicator is cumulative and considering historical persons.

QS forms its rankings, considering such indicators as research activities, teaching, number of foreign students, etc. To be included in the ranking, a university should offer bachelor and post-diploma programs in at least two wide spheres. For example, business and law QS offers a huge variety of rankings, they can be separated by location, program, subjects etc. We are going to concentrate on two of them: QS World University Rankings and QS Graduate Employability Rankings. The first uses the following 6 metrics in its methodological framework: Academic Reputation (40%), Employer Reputation (10%), Faculty/Student Ratio (20%), Citations per faculty (20%), International Faculty Ratio (5%), International Student Ratio (5%). Graduates play an active role in forming the Employer Reputation coefficient, as it shows whether they are prepared good enough for labor market competition. The coefficient is based on the QS Employer Survey, which includes answers of 40,000 responses about identification of institutions from which they source the most competent, innovative, effective graduates. The second one takes into account employer-student connections, alumni outcomes, employer reputation, partnerships with employers and graduate employment rate. QS Graduate Employability Rankings provides information about graduate employability outcomes and perspectives, which is achieved by comparison of different universities. The following indicators were used to form the ranking: Employer Reputation (30%), Partnerships with Employers (25%), Alumni outcomes (20%), Employers' presence on campus (20%), Graduate employment rate (20%). Alumni outcomes are totally based on the success of individuals, as it helps to find out which universities supply labor market with the most wealthy and prosperous students. Graduate employment rate considers the success of students, measuring the number of alumni, who have found a job during the first 12 months after their graduation.

Thus, in the paper we observe the related to alumni scales and the final universities rankings for the 2017 year from QS, QS graduate, ARWU and compare them with the ranking based on the Wikipedia.

## Results

### 3.1 Descriptive statistics

First of all, we prepared descriptive statistics on alumni page-views using public API Wikipedia. We received information about 464 universities and 265,709 alumni. The full dataset contains data about alumni since the X century. Despite the fact that the full dataset about them was gathered, it was decided to concentrate on the XX century in order to trace the alumni influence on universities and on modern society as a whole. As a consequence, additional parameters were entered – "born after 1947" and "has more than 999 views" – specifically to clarify the quality of the modern educational programs (Table 1).

**Table 1** Descriptive statistics of the data about alumni on Wikipedia: number of alumni and number of universities in the dataset depending on the period of their years of birth.

| Dataset | Number of alumni | Number of universities |
|---|---|---|
| Full dataset | 265,709 | 464 |
| 1900-2000 | 190,885 | 463 |
| 1948-2000 | 105,112 | 460 |
| 1948-2000 & >999 views | 88,242 | 460 |

We have made a calculation of mean, median and standard deviation of the page-views numbers (Table 2). According to the results, the gap between the most popular alumni and ranking outsiders increases because of the narrowing of the alumni birth range that enhances the correlation of page-views sum and the page-views standard deviation. It provides evidence that more and more graduates born after 1900 or after 1947 have been introduced on Wikipedia. This conclusion provides wider coverage of alumni in contrast to counting of alumni as Nobel laureates & Fields Medalists, for instance, and gives a more realistic view on alumnus as the results of the universities' production activities.

**Table 2** Descriptive statistics of the data about alumni on Wikipedia: the characteristics of the page-views number depending on the period of their years of birth.

| Dataset | Mean page-views | Median page-views | Standard deviation page-views |
| --- | --- | --- | --- |
| Full dataset | 21,146 | 2,133 | 162,485 |
| 1900-2000 | 25,301 | 2,472 | 180,412 |
| 1948-2000 | 31,778 | 2,983 | 208,063 |
| 1948-2000 & >999 views | 37,728 | 4,025 | 226,596 |

Moreover, contemporary graduates attract more page-views than older university mates regardless of their recognition (Fig. 1).

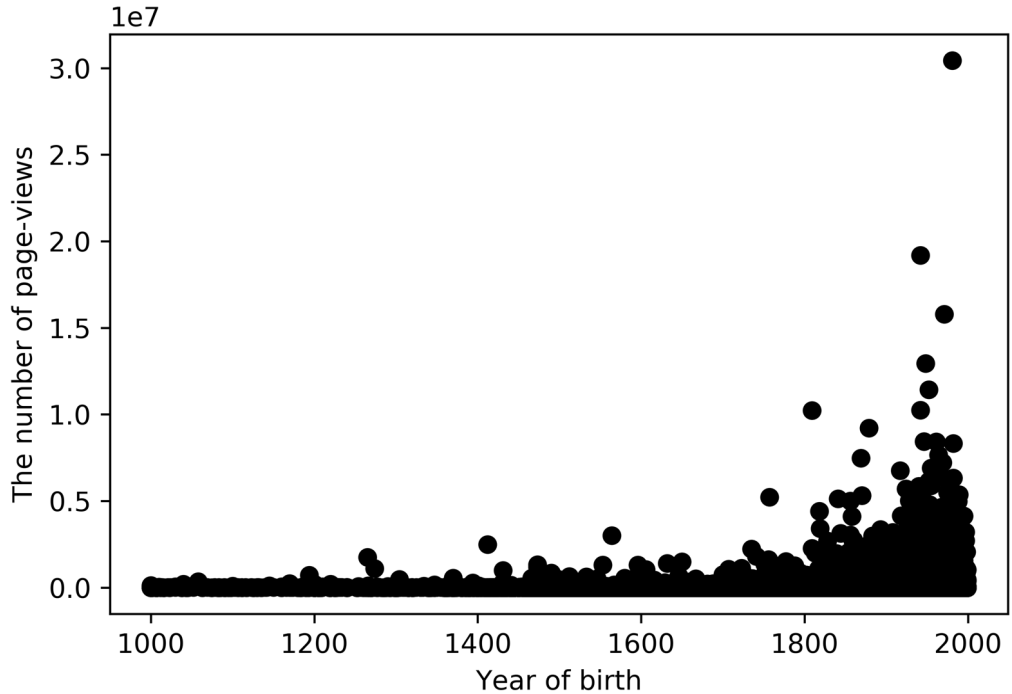

**Fig. 1.** The number of page-views dependence on the year of the graduate birth.

Aggregating the obtain dataset by universities, one can able to construct the ranking based on the sum of page-views of alumni. Further, we will provide the comparison between Wikipedia ranking with the famous world universities ranking. But now we illustrate the correlation between rankings which were created for different periods of year of alumni birth. We divided years of birth on five categories: born anytime, born in XX century, born after 1947, 1965, 1980 (Table 3).

**Table 3** Correlation matrix between rankings based on Wikipedia alumni page-views for different periods of year of alumni birth.

|  | born any time | born in XX century | born after 1947 | born after 1965 | born after 1980 |
|---|---|---|---|---|---|
| born any time | 1.00 |  |  |  |  |
| born in XX century | 0.99 | 1.00 |  |  |  |
| born after 1947 | 0.95 | 0.97 | 1.00 |  |  |
| born after 1965 | 0.89 | 0.92 | 0.96 | 1.00 | |
| born after 1980 | 0.81 | 0.84 | 0.88 | 0.93 | 1.00 |

We can notice from Table 3 that ranking is changed with increasing of year of alumni birth. However, all correlation coefficients are positive and high.

## 3.2   Top-10 graduates based on the popularity on Wikipedia

The world news and events indicate the seasonal popularity of the alumni. For example, Table 4 represents the top-10 graduates in the dataset, Table 5 represents the top-10 graduates that were born after 1947. Some of them have been a long time famous, but the others are famous due to events in the observed time.

**Table 4** Top-10 graduates based on the number of page-views on Wikipedia

| University | Name | Birth year | Views |
|---|---|---|---|
| Northwestern University | Meghan Markle | 1981 | 30,430,581 |
| University of Cambridge | Stephen Hawking | 1942 | 19,183,278 |
| University of Pennsylvania | Elon Musk | 1971 | 15,791,090 |
| University of Cambridge | Charles, Prince of Wales | 1948 | 12,944,420 |
| Saint Petersburg State University | Vladimir Putin | 1952 | 11,426,497 |
| University of Edinburgh | Charles Darwin | 1809 | 10,225,649 |
| ETH Zurich – Swiss Federal Institute of Technology Zurich | Albert Einstein | 1879 | 9,206,148 |
| University of Miami | Sylvester Stallone | 1946 | 8,444,153 |
| Columbia University | Barack Obama | 1961 | 8,418,653 |
| University of St Andrews | Prince William, Duke of Cambridge | 1982 | 8,331,379 |

**Table 5** Top-10 graduates that were born after 1947 based on the number of page-views on Wikipedia

| University | Name | Birth year | Views |
|---|---|---|---|
| Northwestern University | Meghan Markle | 1981 | 30,430,581 |
| University of Pennsylvania | Elon Musk | 1971 | 15,791,090 |
| University of Cambridge | Charles, Prince of Wales | 1948 | 12,944,420 |
| Saint Petersburg State University | Vladimir Putin | 1952 | 11,4264,97 |
| Columbia University | Barack Obama | 1961 | 8,418,653 |
| University of St Andrews | Prince William, Duke of Cambridge | 1982 | 8,331,379 |
| Princeton University | Jeff Bezos | 1964 | 7,661,172 |
| Reed College | Steve Jobs | 1955 | 6,899,464 |
| University of St Andrews | Catherine, Duchess of Cambridge | 1982 | 6,336,213 |
| Harvard University | Bill Gates | 1955 | 5,880,135 |

## 3.3 University rankings

Using described method, the universities ranking based on the alumni from Wikipedia was created and its results were compared with some famous universities rankings: QS, QS graduate, and ARWU (Table 6). These universities rankings were chosen because they contain the scales connected with alumni (see Methods).

**Table 6** Top-10 universities based on different rankings and metrics

| Wikipedia: full dataset | Wikipedia: 1948-2000 & >999 views | ARWU | QS | QS Graduate |
|---|---|---|---|---|
| Harvard University | New York University | Harvard University | Massachusetts Institute of Technology (MIT) | Stanford University |
| New York University | Harvard University | University of Cambridge | Stanford University | University of Oxford |
| Columbia University | University of California, Los Angeles | Massachusetts Institute of Technology (MIT) | Harvard University | The University of Tokyo |
| University of California, Los Angeles | Columbia University | University of California, Berkeley | University of Cambridge | Lomonosov Moscow State University |
| Yale University | University of Southern California | Columbia University | University of Oxford | Massachusetts Institute of Technology (MIT) |
| University of Southern California | Waseda University | University of Chicago | Imperial College London | University of Cambridge |
| Waseda University | Stanford University | Princeton University | National University of Singapore (NUS) | University of California, Los Angeles (UCLA) |
| University of Tokyo | Yale University | University of Oxford | London School of Economics and Political Science (LSE) | University of Pennsylvania |
| Stanford University | Keio University | California Institute of Technology | CentraleSupélec | Columbia University |
| Keio University | Northwestern University | Ecole Normale Superieure - Paris | London Business School | Peking University |

The different contents of the top 10 universities rankings are explained by various parameters included in the assessment of universities where alumni are only one of the criteria. However, the following universities are intersected in the rankings:

- Columbia University,
- Stanford University,

The emergence of these universities in all mentioned rankings indicates not only the high level of the scientific research criteria, but also their importance for society provided by their graduates.

The difference between full dataset Wikipedia ranking and the one after 1947 shows that the universities rarely maintain the tendency of graduating the most influential people. Some of them had the famous alumni in the past; the other has the popular alumni nowadays. Also, we detect the distinction between Wikipedia ranking and the international rankings (QS, ARWU) (Table 6): e.g. in the top 10 universities based on the Wikipedia ranking we observe three Japanese universities which are not represented in the top 10 universities according to famous rankings. This effect can be explained through the national differences in the Web consumption.

Moreover, rankings' correlation was calculated on the base of only those scales, which were connected with alumni (Table 7). The biggest correlation was found between different modifications of the ranking, formed from Wikipedia dataset. The smallest correlation appeared to be between QS and ARWU. It can also be noticed, that the Wikipedia ranking correlates with ARWU more than with QS, however, the correlation coefficient between Wikipedia ranking and QS graduate is close to the correlation coefficient between Wikipedia ranking and ARWU. It may be partly explained by the similarity of these rankings: both take into account famous graduates, who are honored, and, thus, have got a scientific reward (ARWU) or a page in Wikipedia.

**Table 7** Correlation between rankings, based on the scales connected with alumni

|  | Wikipedia: full dataset | Wikipedia: 1900-2000 | Wikipedia: 1948-2000 & >999 views | ARWU | QS | QS Graduate |
|---|---|---|---|---|---|---|
| Wikipedia: full dataset | 1.00 | | | | | |
| Wikipedia: 1900-2000 | 0.99 | 1.00 | | | | |
| Wikipedia: 1948-2000 & >999 views | 0.94 | 0.97 | 1.00 | | | |
| ARWU | 0.52 | 0.49 | 0.42 | 1.00 | | |
| QS | 0.35 | 0.35 | 0.36 | 0.28 | 1.00 | |
| QS Graduate | 0.48 | 0.46 | 0.42 | 0.38 | 0.63 | 1.00 |

## 3.4 Webpages of universities on Wikipedia.

To compare the popularity of the universities' webpages on Wikipedia with the popularity of their alumni there, we have made the two separate rankings (Table 8). The correlation of the number of the university page-views and the university graduates' views is equal to 0,83. This indicates that the popularity of graduates reflects the popularity of the university. Unfortunately, we cannot prove now that the alumni popularity leads to the increase in number of page-views of the universities webpages, but logically Wikipedia users follow the link on university from the alumni webpages.

**Table 8** Top 10 universities based on the alumni and universities popularity on Wikipedia.

| Wikipedia: full dataset | Wikipedia: 1948-2000 & >999 views | Wikipedia: webpages of universities |
|---|---|---|
| Harvard University | New York University | Harvard University |
| New York University | Harvard University | University of Oxford |
| Columbia University | University of California, Los Angeles | Stanford University |
| University of California, Los Angeles | Columbia University | Columbia University |
| Yale University | University of Southern California | Massachusetts Institute of Technology |
| University of Southern California | Waseda University | Yale University |
| Waseda University | Stanford University | University of Cambridge |
| University of Tokyo | Yale University | Princeton University |
| Stanford University | Keio University | University of California, Berkeley |
| Keio University | Northwestern University | University of Pennsylvania |

## Limitations

The method has the following limitations that should be taken into consideration. For instance, there is a close dependence on the influence of certain current events or news. However, the method allows observing the dynamics of the alumni popularity. Next, it is necessary to take into consideration the year of the university's foundation, the target set of the students, the specifics of the university, the alumni role in society (e.g. president, singer or scientist). Finally, the proposed algorithm can be improved to increase the accuracy of the dataset.

## Conclusion

In this study, we discuss the possibility to evaluate the university contribution to society. We focus on investigating the university impact based on popularity (number of page-views) of their alumni's pages on Wikipedia. This method is also aimed to provide wider coverage of alumni in research. It deals with the information crawling from a web-based free encyclopedia Wikipedia. We used data from Wikipedia: all alumni from 464 top universities were found and the number of page-views (of their webpages on Wikipedia) was calculated. We propose to evaluate the influence of the alumni on the society through the popularity of their webpages on Wikipedia. The perspective of this approach is proved by the higher number of the contemporary persons page-views in comparison to great people of the past. This method reduces the limitation of the current universities ranking in graduate research: to take into consideration alumni who influence the whole society, not only in business or science. The method allows to define the contribution in society in past or to track every-year changes in popularity based on the alumni page-views. For potential students at universities, it may be interesting to know the success of recent graduates. For example, QS provides some metrics to evaluate graduates efficiently at work that is highly important during the period of choosing between universities (QS Graduate Employability Rankings 2019). However, the popularity of alumni on Wikipedia as well as the possibility to get acquainted with the history of their success give less abstract information rather than percentiles in employee's rankings.

In this paper, we have created the universities ranking based on the alumni and compared it with the top three universities rankings which are included alumni analyses in the scales: QS, ARWU, and QS Graduate Employability Rankings. The results of this comparison illustrate a connection between these indicators that support the adequacy of the Wikipedia approach. However, we suggest considering our method as an addition to the already created methodology.

Finally, we compared the alumni ranking with the ranking based on the popularity of the universities' webpages on Wikipedia: there is a strong connection between them (0,83). So, we observed the connection between alumni and universities. We assume that the popularity of graduates reflects the popularity of the university.

Although the method is not completely free from limitations now, it has been used quite successfully in this study. At the same time, we will be working on the improvement of the method. Additional information about alumnus achievements in particular of activities spheres can help to deepen the research. Addition to the existing metrics in the university rankings, the contribution of the alumni to the society can make the studies more powerful, precise and complete. Further investigation will shed light on the open question of the university impact.

**Funding:** This work was supported by The Tomsk State University competitiveness improvement program and by the Basic Research Program of the National Research University Higher School of Economics.

# References

Agasisti, T. (2017). Management of Higher Education Institutions and the Evaluation of their Efficiency and Performance. *Tertiary Education and Management*, 23(3), 187–190. doi:10.1080/13583883.2017.1336250

Altmetric (2019). What are Altmetrics? https://www.altmetric.com/about-altmetrics/what-are-altmetrics/ Accessed 26 April 2020.

Brooks, R. (2005). Measuring university quality. *The Review of Higher Education*, 29(1), 1–21.

Cabrera, A. F., Weerts, D. J., & Zulick, B. J. (2005). Making an impact with alumni surveys. New Directions for Institutional Research, 2005(126), 5–17.

Chavda, J., & Patel, A. (2016). Measuring research impact: bibliometrics, social media, altmetrics, and the BJGP. *Br J Gen Pract,* 66(642), e59–e61.

Cole, L., & Thompson, G. (2002). Satisfaction of agri-business employers with college graduates they have hired. *NACTA Journal*, 34–39.

Fumasoli, T. (2016). The Roles of the University in Society. *Higher Education Quarterly*, 70(2), 106–107.

Gault, J., Redington, J., & Schlager, T. (2000). Undergraduate business internships and career success: are they related? *Journal of marketing education*, 22(1), 45–53.

Gonçalves, G. R., Ferreira, A. A., de Assis, G. T., & Tavares, A. I. (2014). Gathering alumni information from a web social network. *In 2014 9th Latin American Web Congress* (pp. 100–108). IEEE.

Hernández-March, J., Del Peso, M. M., & Leguey, S. (2009). Graduates' skills and higher education: The employers' perspective. Tertiary education and management, 15(1), 1–16.

Kempe, D., Kleinberg, J., & Tardos, É. (2005). Influential nodes in a diffusion model for social

networks. *In International Colloquium on Automata, Languages, and Programming* (pp. 1127-1138). Springer, Berlin, Heidelberg.

Khazragui, H., & Hudson, J. (2014). Measuring the benefits of university research: Impact and the REF in the UK. *Research Evaluation*, 24(1), 51–62.

Kosmützky, A., & Ewen, A. (2016). Global, national and local? The multilayered spatial ties of universities to society. *In RE-BECOMING UNIVERSITIES?* (pp. 223–245). Springer.

Kozitsin, I. V. (2020). Formal models of opinion formation and their application to real data: evidence from online social networks. The Journal of Mathematical Sociology, 1-28.

Kozitsin, I. V., & Chkhartishvili, A. G. (2020, September). Users' Activity in Online Social Networks and the Formation of Echo Chambers. In 2020 13th International Conference" Management of large-scale system development"(MLSD) (pp. 1-5). IEEE.

Laredo, P. (2007). Revisiting the third mission of universities: toward a renewed categorization of university activities? *Higher education policy*, 20(4), 441–456.

Laschinger, H. K. S. (2012). Job and career satisfaction and turnover intentions of newly graduated nurses. *Journal of nursing management*, 20(4), 472–484.

Lee, R., & Wilbur, E. R. (1985). Age, education, job tenure, salary, job characteristics, and job satisfaction: A multivariate analysis. *Human Relations*, 38(8), 781–791.

Linkedin (2019). www.linkedin.com Accessed 26 April 2020.

Lukman, R., Krajnc, D., & Glavič, P. (2010). University ranking using research, educational and environmental indicators. *Journal of Cleaner Production*, 18(7), 619–628.

Montesinos, P., Carot, J. M., Martinez, J., & Mora, F. (2008). Third mission ranking for world class universities: Beyond teaching and research. *Higher education in Europe*, 33(2–3), 259–271.

Nann, S., Krauss, J. S., Schober, M., Gloor, P. A., Fischbach, K., & Führes, H. (2010). The power of alumni networks-success of startup companies correlates with online social network structure of its founders.

https://dspace.mit.edu/bitstream/handle/1721.1/66586/CCI-2010-001.pdf?...1 Accessed 26 April 2020.

Nariani, R. (2017). Supplementing Traditional Ways of Measuring Scholarly Impact: The Altmetrics Way. https://yorkspace.library.yorku.ca/xmlui/bitstream/handle/10315/33652/SupplementingTraditionalWays_RN.pdf?sequence=1 Accessed 26 April 2020.

Ng, T. W. H., Eby, L. T., Sorensen, K. L., & Feldman, D. C. (2005). Predictors of objective and subjective career success: A meta-analysis. *Personnel psychology*, 58(2), 367–408.

Nicolescu, L., & Paun, C. (2009). Relating Higher Education with the Labour Market: Graduates' expectations and employers' requirements. *Tertiary education and management*, 15(1), 17–33.

Olcay, G. A., & Bulu, M. (2017). Is measuring the knowledge creation of universities possible?: A review of university rankings. *Technological Forecasting and Social Change*, 123, 153–160.

Papoutsoglou, M., Mittas, N., & Angelis, L. (2017). Mining People Analytics from StackOverflow Job Advertisements. *In 2017 43rd Euromicro Conference on Software Engineering and Advanced Applications* (SEAA) (pp. 108–115). IEEE.

QS Graduate Employability Rankings (2017). https://www.topuniversities.com/university-rankings/employability-rankings/2017 Accessed 26 April 2020.

QS Graduate Employability Rankings (2019). https://www.topuniversities.com/university-rankings/employability-rankings/2019 Accessed 26 April 2020.

Quarmby, K. L., Willett, P., & Wood, F. E. (1999). Follow-up study of graduates from the MSc Information Management programme at the University of Sheffield. *Journal of information science*, 25(2), 147–155.

Ranking Web of Universities (2019). http://www.webometrics.info/en/Europe Accessed 26 April 2020.

Rubens, N., Russell, M., Perez, R., Huhtamäki, J., Still, K., Kaplan, D., & Okamoto, T. (2011). Alumni network analysis. *In 2011 IEEE Global Engineering Education Conference (EDUCON)* (pp. 606–611). IEEE.

Sánchez-Barrioluengo, M. (2014). Articulating the 'three-missions' in Spanish universities. *Research Policy*, 43(10), 1760–1773.

Scillitoe, J. L. (2013). The role of alumni attachment in the university technology transfer process. *In 2013 Proceedings of PICMET'13: Technology Management in the IT-Driven Services* (PICMET) (pp. 2397–2406). IEEE.

Scott, J. C. (2006). The mission of the university: Medieval to postmodern transformations. *The journal of higher education*, 77(1), 1–39.

Secundo, G., Perez, S. E., Martinaitis, Ž., & Leitner, K. H. (2017). An Intellectual Capital framework to measure universities' third mission activities. *Technological Forecasting and Social Change*, 123, 229–239.

Seibert, S. E., Crant, J. M., & Kraimer, M. L. (1999). Proactive personality and career success. *Journal of applied psychology*, 84(3), 416.

Shanghai Ranking (2017). http://www.shanghairanking.com/ARWU2017.html Accessed 26 April 2020.

Stack Overflow (2019). https://www.stackoverflowbusiness.com/talent/platform/source/job-listings Accessed 26 April 2020.

The World University Rankings (2019). https://www.timeshighereducation.com/ Accessed 26 April 2020.

Top Universities (2017). https://www.topuniversities.com/university-rankings/world-university-rankings/2016 Accessed 26 April 2020.

Uyarra, E. (2010). Conceptualizing the regional roles of universities, implications and contradictions. *European Planning Studies*, 18(8), 1227–1246.

Vermeulen, L., & Schmidt, H. G. (2008). Learning environment, learning process, academic outcomes and career success of university graduates. *Studies in Higher Education*, 33(4), 431–451. doi:10.1080/03075070802211810

Wikipedia (2019). https://en.wikipedia.org/wiki/Wikipedia Accessed 26 April 2020.

Wikipedia:Notability (people) (2019). https://en.wikipedia.org/wiki/Wikipedia:Notability_(people) Accessed 26 April 2020.

Witte, K. De, & López-Torres, L. (2017). Efficiency in education: a review of literature and a way forward. *Journal of the Operational Research Society*, 68(4), 339–363.

# Appendix

The description of the method to crawl information about alumni and universities from a web-based free encyclopedia Wikipedia.

Firstly, we created the database which contained all the webpages of the chosen universities on Wikipedia where all the versions of the redirect pages (to the universities) in English and national languages were mentioned. Secondly, we identified all the webpages devoted to people. Thus, a complete copy of all Wikipedia wikis (later — dump) (in the form of XML text) was obtained (relevant for September 1, 2018, in 18 languages). Using the list of universities, we programmatically composed the table which includes university's Id; the name of the webpage on Wikipedia; language domain (ru, en, fr, etc) with the all direct pages and the webpages in English and/or national languages.

For the dump we performed the following algorithm:

- If the file has not been completed the following independent webpage inside the dump is read.

- In the webpage, depending on the language used, the keywords from the special dictionary are found. For example, the following phrases are included in the dictionary: "born", "Category: 1980 births".

- If the markers of a person's page are found, then a birth year is found (the first four numbers at the beginning of the page, no further than in 1000 words from the start). If the birth year is not found in the first 1000 words (because of the traditional article structure on Wikipedia), it stays empty.

- The page is saved in the independent file .xml with the name "page_Id_birthday".

We performed the following algorithm for all files from the previous step:

- If the file has not been completed the next sentence is read (all words until a full stop). The entire link to the other Wiki-pages is saved in the database.

- In the obtained sentence the trigger-words from the definite set are searched. Under a definite set we mean the words indicating the connection with the university ("graduated", "alumni", "received degree", etc. (case-insensitive)). Searched words and collocations were prepared by translators on the basis of manual sampling from the webpages of graduates in 18 languages.
- If such words are found, then for every link in the obtained database a match with the universities' names is searched.
- If the match is found then in a final file the following row is added: "University's Id", "University's name", "link to the webpage of graduates", "birth year", "language domain".
- The next file is opened.

For each record with the language domain different from "en", the existence of the webpage in the English domain for this person is checked with help of public API Wikipedia. If the webpage exists so the link to this page and the language domain ("en") are added to the record. To calculate the number of views the obtained table was read by rows. The webpages in English and national languages were extracted. Using public API Wikipedia, we calculated the statistics of the total amount of views for the last year. Values for the webpage in English and national language were summarized and added to the table. To find statistics of page-views of the universities' webpages on Wikipedia, the following method was performed. Using public API Wikipedia, we calculated the statistics of the total amount of views for the last (2017) year — the sum of page-views in English and national languages. Names of all universities were translated into national languages. These algorithms were launched and implemented on September 1st, 2018. The estimated accuracy was found. A selective manual check of the results was carried out (the final dataset contained 6-8% of mistakes).